%Paper: gr-qc/9210017
%From: wms@sze.wustl.edu (Wai Suen)
%Date: Wed, 28 Oct 92 17:12:54 CST

\hoffset=.25in
\vbadness=10000
%
% The following statements redefine the basic fonts
% to be magnified by a factor 1.2
%

\font\bf=cmbx10 scaled 1200

\font\it=cmti10 scaled 1200
\font\sl=cmsl10 scaled 1200

\font\tenrm=cmr10 scaled 1200
\font\sevenrm=cmr9
\font\fiverm=cmr7
\font\teni=cmmi10 scaled 1200
\font\seveni=cmmi9
\font\fivei=cmmi7
\font\tensy=cmsy10 scaled 1200
\font\sevensy=cmsy9
\font\fivesy=cmsy7

\font\tenbf=cmbx10 scaled 1200
\font\sevenbf=cmbx7 scaled 1200
\font\fivebf=cmbx5 scaled 1200
\font\tensl=cmsl10 scaled 1200
\font\tentt=cmtt10 scaled 1200
\font\tenit=cmti10 scaled 1200
\catcode`\@=11
\textfont0=\tenrm \scriptfont0=\sevenrm \scriptscriptfont0=\fiverm
\def\rm{\fam\z@\tenrm}
\textfont1=\teni \scriptfont1=\seveni \scriptscriptfont1=\fivei
\def\mit{\fam\@ne} \def\oldstyle{\fam\@ne\teni}
\textfont2=\tensy \scriptfont2=\sevensy \scriptscriptfont2=\fivesy
\def\cal{\fam\tw@}
\textfont3=\tenex \scriptfont3=\tenex \scriptscriptfont3=\tenex
\newfam\itfam \def\it{\fam\itfam\tenit} % \it is family 4
\textfont\itfam=\tenit
\newfam\slfam \def\sl{\fam\slfam\tensl} % \sl is family 5
\textfont\slfam=\tensl
\newfam\bffam \def\bf{\fam\bffam\tenbf} % \bf is family 6
\textfont\bffam=\tenbf \scriptfont\bffam=\sevenbf
\scriptscriptfont\bffam=\fivebf
\newfam\ttfam  % \tt is family 7
\textfont\ttfam=\tentt
\catcode`\@=12
%
% This ends font redefinitions
%
\rm
\hfuzz=10pt \overfullrule=0pt
\vsize 8.75in
\hsize 6in
\def\doublespace{\baselineskip=28pt}

\parindent 20pt \parskip 6pt
\def\blankline{\par\vskip \baselineskip}

\def\lbbrack{{\lbrack\!\lbrack}}
\def\rbbrack{{\rbrack\!\rbrack}}
\doublespace
\parindent=1cm
\raggedbottom
\centerline{\bf Is Quantum Spacetime Foam Unstable?}
\blankline
\centerline{Ian H.~Redmount}
\centerline{\sl Department of Physics, University of Wisconsin--Milwaukee}
\centerline{\sl Milwaukee, Wisconsin~~~53201}
\centerline{and}
\centerline{Wai-Mo Suen}
\centerline{\sl Department of Physics, Washington University}
\centerline{\sl St.~Louis, Missouri~~~63130--4899}
\blankline
\centerline{(Submitted to {\it Physical Review Letters\/} 26 June 1992}
\blankline
\centerline{\bf ABSTRACT}\par\nobreak
A very simple wormhole geometry is considered as a model of a mode of
topological fluctutation in Planck-scale spacetime foam.  Quantum dynamics
of the hole reduces to quantum mechanics of one variable, throat radius, and
admits a WKB analysis.  The hole is quantum-mechanically unstable:  It has
no bound states.  Wormhole wave functions must eventually leak to large
radii.  This suggests that stability considerations along these lines may
place strong constraints on the nature and even the existence of
spacetime foam.
\vfil\eject
\parskip=14pt
\centerline{\bf I.~~INTRODUCTION}\par\nobreak
Some 35 years ago Wheeler~[1] made a remarkable suggestion, based on
dimensional arguments:  On Planck-length scales spacetime fluctuates
quantum-\discretionary{mechani-}{cally}{mechanically}, so randomly and
violently that it develops all kinds of microscopic topological structures,
such as ``wormholes,'' although on larger scales it appears smooth and simply
connected.  It is distressing that after so many years our knowledge of quantum
gravity is still far from being able to confirm or disprove the existence of
this ``spacetime foam.''

Obstacles to analyzing this conjecture are apparent.  It is well known~[2]
that a Lorentzian manifold must become singular or degenerate at points of
topological change, or admit closed timelike paths.  Formulation of field
theory on such a manifold is plainly problematic.  Such difficulties might
be avoided by treating Euclidean manifolds~[3]; unlike the Euclideanization
of ordinary field theory, equivalent to the Lorentzian formulation in the
sense of contour integration, Euclidean quantum gravity is physically
different from Lorentzian.  Indeed, maybe spacetime is intrinsically
Euclidean on the Planck scale, characterized by Euclidean quantum foam,
and $3+1$-dimensional Lorentzian spacetime only emerges after a transition
to a classical regime.  However, Euclidean quantum gravity has fundamental
difficulties of its own~[4], most notably failure of the Euclidean action
to be positive definite, the problem of interpretation, and recovery of
Lorentzian spacetime.

Both Euclidean and Lorentzian versions of quantum foam have received much
recent attention.  The focus on the Euclidean version is on its possible role
in determining fundamental constants~[5], while for the Lorentzian version it
has been suggested that a microscopic wormhole might be extracted from the
foam to produce a traversable macroscopic wormhole or a time-machine~[6].
But though both sets of ideas are ingenious and have far-reaching consequences
for other areas of physics, neither sheds much light on the actual existence or
structure of the foam itself.

Here we present a simple analysis probing the stability of spacetime foam,
to examine the constraints placed on its existence and structure by the
apparent absence of topological structure to spacetime on macroscopic scales.
We picture Lorentzian spacetime filled with many different sorts of microscopic
wormholes, fluctuating into existence, living for microscopic time periods,
and pinching off.  At moments of birth and pinch-off of holes Euclideanization
may or may not be needed; we do not treat the actual points of topological
change here.

Some of these structures are easily modeled classically:  Wormholes
can be constructed by excising a ``world tube'' from some $3+1$-dimensional
spacetime and joining this to another such spacetime, with a corresponding
excision~[7].  These are extreme versions of situations in which the
curvature in a wormhole throat is greater than that of the surrounding
spacetime; here the curvature at the join or throat is a delta-function
distribution.  This approximation is very useful for simplifying the
dynamics.  The stress-energy at the throat is determined by the Einstein
field equations in the form of junction conditions~[8].  This necessarily
violates the weak energy condition, with negative energy density (in some
reference frames) somewhere in the throat.  This is not itself a fatal
flaw of the wormholes~[6,9]---while it might help account for the absence
of macroscopic holes, it does not rule out the possibility of quantum,
Planck-scale ones, nor does it guarantee that microscopic holes would
not grow in size.

To make the analysis tractable we treat the simplest such wormhole:  that
obtained by excising a spherical region, with time-varying radius, from two
Minkowski spaces and joining them~[7].  The quantum-gravitational dynamics
of the model reduces to quantum mechanics of a single degree of freedom, the
throat radius, yielding a ``minisuperspace model'' for spacetime foam.  The
quantum wormhole is described by a wave function depending on that radius and
time, as defined in the external flat spaces.  If the wave function is
localized about some Planck-scale radius at some initial time, what will
be its subsequent evolution?

There is no standard approach to the quantization of a system like this.
Evolution in a time coordinate defined by the Minkowski spaces external to
the wormhole is at issue.  Hence the familiar Dirac quantization procedure,
giving rise to a time-independent wave function solving the Wheeler-de~Witt
equation~[10], is not suitable.  Instead we impose the Hamiltonian constraint
classically, using it to reduce the phase space of the system. We construct
an action for the dynamics in the reduced phase space, and quantize the
system by using this action in a Feynman path integral.  The resulting
propagator for wormhole wave functions is evaluated in a WKB approximation.

The results indicate that these wormholes are quantum-mechanically unstable:
Though the classical evolution of the throat radius may be bounded, the
quantum propagator admits no decomposition into contributions from any
spectrum of bound and continuum states.  Rather its behavior is akin to
that of a ``leaking'' system, such as a particle confined by finite walls.
This implies that wormhole wave functions must eventually ``leak'' to
arbitrarily large throat-radius values.  (Such quantum instability of a
classically stable object is familiar, as in the particle case.  So too in
a gravitational context:  Classically stable black holes are subject to
Hawking evaporation.)  The wormholes thus suggest a possible unstable
mode of spacetime foam, microscopic topological structure growing eventually
to macroscopic size.  Numerical calculations of wave-function evolution
show that the time scale of this instability might be very long, in
terms of the Planck scales appropriate to the model---though perhaps not
on scales of observational significance.

This simple analysis thus points up a line of inquiry potentially of great
significance.  If more detailed, comprehensive analyses substantiate the
existence of an unstable mode, then that together with the observed absence
of macroscopic wormholes might indicate that spacetime does not possess
microscopic topological structure---of Lorentzian signature---after all.
Lorentzian spacetime foam could be inconsistent with known gravitational
and quantum theory and observation.\par
\blankline
\centerline{\bf II.~~WORMHOLE QUANTUM MECHANICS}\par\nobreak
A classical, spherically symmetric ``Minkowski wormhole''~[7] is
constructed by:  excising a sphere of radius $r=R(t)$, with~$R$ some
function of a Minkowski time coordinate~$t$, from two copies of Minkowski
spacetime; identifying the two boundary surfaces~$r=R(t)$; and incorporating
an appropriate surface-layer stress-energy on the boundary to satisfy
the Einstein field equations.  Off the boundary both exterior spacetime
regions are flat and empty, so the field equations are satisfied trivially.
On the boundary---now the throat of the wormhole---the Einstein equations
are equivalent to the junction conditions~[8]
$$S^i_j={1\over8\pi}\,\lbbrack K^i_j-\delta^i_j\,K^m_m\rbbrack\ ,
\eqno(1)$$
where $S^i_j$ is the surface stress-energy tensor and the right-hand
side is the discontinuity in the extrinsic curvature~$K^i_j$, minus
its trace~$K^m_m$, across the boundary.  (Units with $G=1$, as well
as $\hbar=c=1$, are used throughout.)  For this wormhole geometry the
junction conditions take the form
$$\eqalignno{S_{\tau\tau}&=-{1\over2\pi R}\,{1\over(1-\dot{R}^2)^{1/2}}
&(2{\rm a})\cr
\noalign{\hbox{and}}
S_{\theta\theta}&={1\over4\pi}\left({R\over(1-\dot{R}^2)^{1/2}}+
{R^2\ddot{R}\over(1-\dot{R}^2)^{3/2}}\right)\ ,&(2{\rm b})\cr}$$
where overdots denote derivatives with respect to Minkowski-coordinate
time~$t$ (in a frame in which the boundary sphere expands or contracts
but does not translate), and the boundary coordinates~$\tau$ and~$\theta$
are proper time---related to coordinate time via
$d\tau=(1-\dot{R}^2)^{1/2}dt$---and polar angle, respectively.  These give
the classical equation of motion for the wormhole, once an equation of state
relating the surface density $\sigma=S_{\tau\tau}$ and pressure
$p=S_{\theta\theta}/R^2$ of the matter on the throat is specified.

The equation of state could be chosen to make the equation of motion
simple.  For example, the choice $p=-\sigma/2$ would imply $\ddot{R}=0$.
The quantization of the system thus described is trivial:  The wave function
evolves as that of a free particle.  A wave function initially concentrated
about some $R$~value will disperse to infinity.  However, we do not expect
such a wormhole, which evolves classically with its throat radius either fixed,
or expanding or collapsing linearly, to correspond to those fluctuating into
existence in spacetime foam.

Instead we choose an equation of state such that the equation of motion
describes expansion from zero radius to some maximum value and recollapse.
The classical behavior of the model thus accords with that, e.g., of a
Schwarzschild wormhole, and that expected of a foam-like fluctuation.
Specifically, we use
$$p=-\sigma/4\ ,\eqno(3)$$
which yields
$$2R\ddot{R}-\dot{R}^2+1=0\ .\eqno(4)$$
The solutions of this equation are parabolic trajectories:
$$R_{\rm cl}(t)={1\over\alpha}\left(1-{\alpha^2(t-t_0)^2\over4}\right)\ ,
\eqno(5)$$
where $\alpha$ and $t_0$ are constants.

The quantum dynamics of the wormhole can be described via a Feynman path
integral.  An action corresponding to Eq.~(4), obtained from the integral
of the scalar curvature of the wormhole geometry, is
$$S=\int\left(R\dot{R}\ln\left|{1+\dot{R}\over1-\dot{R}}\right|-2R\right)\,dt
\ .\eqno(6)$$
Reduced to the single dynamical variable~$R$, the system resembles a point
particle in one dimension, with a complicated ``kinetic term'' in the action.
(In this respect it is similar to a relativistic free particle~[11].)  The
wormhole is described by a wave function~$\psi(R,t)$, the evolution of which
may be given thus:
$$\psi(R,t)=\int G[R,t;R_0,0]\,\psi_0(R_0)\,dR_0\ .\eqno(7)$$
The propagator is given by
$$G[R,t;R_0,0]=\int_C e^{iS[R(t)]}\,{\cal D}[R(t)]\ ,\eqno(8)$$
with $C$ denoting the class of paths included in the path integral.  All paths
moving forward in~$t$, with $R(t)\ge0$, are included.  The latter restriction
can be implemented as for a point particle confined to a half space, i.e.,
as if there were an infinite potential wall at~$R=0$.  This implies the
boundary condition $\psi(0,t)=0$.  By imposing this condition we exclude
consideration of topology-changing processes---wormhole creation or
disappearance---at~$R=0$, but this will not affect our conclusions concerning
the stability of the wormhole.  These follow from the behavior of wave
functions at finite radii, as shown below.

The propagator~(8), with action~(6), can be evaluated approximately.  In the
WKB limit the path integral is dominated by the contributions of classical
paths and small fluctuations about those paths; it takes the form~[12]
$$G[R,t;R_0,0]\sim\sum_{{\rm Classical}\atop{\rm Paths}}\left({i\over2\pi}\,
{\partial^2S[R_{\rm cl}]\over\partial R\partial R_0}\right)^{1/2}\,
e^{iS[R_{\rm cl}]}\ .\eqno(9)$$
The classical paths in the sum include the trajectory of form~(5) between the
initial and final values, plus---owing to the restriction $R\ge0$---paths
between those values which are piecewise of form~(5) but which ``bounce''
one or more times at~$R=0$, the bounce times determined by the requirement
that these paths too be extrema of~$S$.  That condition takes the form of
a cubic equation for the bounce time of a single-bounce trajectory, yielding
one or three such paths, and a quartic equation for the bounce times of
multiple-bounce trajectories, yielding four or two paths with a given number
of bounces up to a maximum number.  Hence the WKB approximation for~$G$ can
be written
$$G^{\rm (WKB)}=\sum_{n=0}^{n_{\rm max}(R_0,R,t)}\sum_kG_n^{(k)}\ ,\eqno(10)$$
where $n$ is the number of bounces, $k$ labels the $n$-bounce paths, and
$G_n^{(k)}$ is the corresponding contribution.  Each of these is of the form
on the right-hand side of Eq.~(9); the prefactors and classical actions
are complicated functions of $R_0$, $R$, $t$, $n$, and the bounce times,
but they can be obtained explicitly in closed form~[13].  The relative phases
of the contributions are determined by the boundary condition $\psi(0,t)=0$.
The propagator is nonvanishing outside the light cone, hence acausal, because
spacelike as well as timelike paths are included in the path integral.  This
is in accord with, e.g., the suggestion of Hartle~[14] that acausal histories
should be included in path integrals for quantum gravity.  It also accords
with the case of the relativistic point particle, for which spacelike paths
must be included in the path integral to obtain agreement with the results
of canonical quantization~[11].

The result reveals the quantum instability of the wormhole.  The
``ground-state energy'' of the hole should follow from the Feynman-Kac~[15]
formula
$$E_0=-\lim_{\tau\to+\infty}{1\over\tau}\ln G[R,-i\tau;R,0]\ .\eqno(11)$$
But the result we obtain for the propagator~(10) indicates that the
$\tau\to\infty$ behavior of~$G$ is
$$G\sim-\left({1\over\pi i}\right)^{1/2}\,\sum_{n=1}^{[\tau/4R]}
{e^{i\tau^2/(4n)}\over n^{1/2}}\ .\eqno(12)$$
Hence the right-hand side of Eq.~(11) does not approach a definite limit:
Its real part tends to zero while its imaginary part oscillates.  This
implies that the wormhole has no spectrum of bound states.
Such behavior is reminiscent of systems, e.g., with ``inverted'' potentials
diverging to negative infinity, or of metastable systems such as a particle
confined by finite walls.  The former case corresponds to rapid growth
of a wormhole to large size; the latter to eventual ``leaking'' to large
size, though the wormhole might remain near its initial size for a long
time.  It is the latter behavior which appears to characterize our model.
The evolution via the propagator~(10) of a wormhole wave function is
illustrated in Fig.~1.  In this example the initial wave
function~$\psi_0(R_0)$ is simply a real Gaussian centered at~$R_0=10$,
with standard deviation~$1/\sqrt2$, all quantities in Planck units.
The wave packet collapses to~$R\approx0$ and rebounds to~$R=10$ to begin
again, following a bouncing classical trajectory piecewise of form~(5).
Its behavior over many oscillations is indicated by the asymptotic behavior
of the propagator~[13]:  In the limit~$t\gg R,R_0$, the largest contributions
to~$G^{\rm (WKB)}$ in Eq.~(10) are certain of the $n=n_{\rm max}$ terms,
which give rise to caustics at intervals corresponding to classical bouncing.
The singularities in the propagator at these points are integrable; they yield
a peak in the wave function which follows a classical trajectory, but with
amplitude decreasing as~$t^{-1/2}$.  The other terms in Eq.~(10) give a
combined contribution to the wave function, at radii near that of the initial
peak, which appears to fluctuate---without dying away---for at least some
hundreds of thousands of classical bounce times.  The wormhole behaves not
unlike an alpha particle, which may oscillate millions of times within a
nucleus before escaping to infinity.\par
\blankline
\centerline{\bf III.~~CONCLUSIONS}\par\nobreak
Spherically symmetric Minkowksi wormholes~[7] provide a very simple model
of a mode of topological fluctuation in Lorentzian spacetime foam, and
suggest a mode unstable against growth to macroscopic size.  The
quantum-gravitational dynamics of these wormholes is reduced to
quantum mechanics of one variable, the throat radius, by describing
the matter at the wormhole throat with a suitable equation of state and
imposing the Hamiltonian constraint classically to reduce the
phase space of the system.  A corresponding reduced action is used in
a Feynman path integral to obtain the propagator for wormhole wave functions;
this is evaluated in the WKB approximation.  The result shows that although
the classical evolution of a wormhole may be bounded, i.e., stable, the hole
nonetheless has no stable bound quantum states, and will eventually grow to
large size by quantum ``diffusion.''

Many systems exhibit similar behavior.  For a particle with the familiar
quadratic kinetic term in the action, the form of the potential
determines whether such diffusion or spreading occurs:  A potential well
with walls falling off at large distances will allow a classically bound
particle to leak out via quantum tunneling, while one which increases
monotonically with distance will not.  For these wormholes, with
more complicated action~(6), so simple an analysis is not possible.  The
more detailed examination of the wormhole propagator described here is
needed to show the instability.

The existence of an unstable mode of fluctuation, such as suggested by
this minisuperspace analysis, would have profound implications.  Since a
macroscopic structure of wormholes is not observed, i.e., spacetime appears
smooth and simply connected on all observable scales, it could indicate that
spacetime does not possess (Lorentzian) foamlike structure on Planck scales.
Whatever features might characterize the quantum behavior of spacetime,
topological structures such as wormholes unstable against growth could
not appear.

The stability of spacetime foam, then, needs more comprehensive study,
to go beyond the limitations of our present calculations.  The most
fundamental of these is our restriction of the gravitational degrees
of freedom to those of the spherically symmetric Minkowski wormhole, i.e.,
the use of a minisuperspace model for topological structure.  In fact
our model is even more restricted than the usual minisuperspace models~[3],
since the matter in the hole is treated not as a dynamical field but via an
equation of state.  Moreover we use the particular equation of state~(3), to
simplify the calculations; other choices give rise to somewhat different
dynamics.\footnote{*}{For example, if the familiar ``dust'' equation of state,
$p=0$, is used, the classical dynamics of the wormhole is only slightly
different:  The equation of motion is $R\ddot{R}-\dot{R}^2+1=0$ instead of
Eq.~(4), and the classical trajectories are sine functions instead of the
parabolas~(5).  The quantum dynamics is amenable to a different treatment
than that described here~[16].}  Also we analyze the model via quantization
in the reduced phase space.  In the absence of a general framework for
quantum-gravity calculations, this method seems best suited to the problem.
It does differ markedly, though, from the Wheeler-de~Witt approach~[3,10].
Here we use the particular reduced action~(6); other forms corresponding to
the classical equation~(4) are possible, leading to different descriptions
of the wormhole's quantum behavior~[16].  Our calculations are carried out in
the WKB approximation.  This is certainly expected to be valid in the late-time
limits in which the instability is manifest.  And WKB calculations of quantum
instabilities in classically stable systems---tunnelling processes, for
example---are well known.  But with no exact solution for comparison it is
difficult to confirm the accuracy of the approximation.  Finally, we implement
the restriction that that throat radii are nonnegative as for a particle in a
half space, with the boundary condition~$\psi(0,t)=0$.  Other implementations
might be used, the most general condition being only that $\psi$ entail no
current in the $-R$~direction at~$R=0$.  Our choice eliminates from
consideration any processes such as wormhole creation or disconnection
at~$R=0$; including these would add an entirely new dimension to the problem,
but it should not alter the instability.  Even with all these assumptions the
calculations are dauntingly difficult~[13].  But it is to be hoped that further
work along these lines will provide valuable insight into the quantum dynamics
of spacetime.\par
\blankline
\centerline{\bf ACKNOWLEDGMENTS}\par\nobreak
We thank J.~L.~Friedman, M.~Visser, C.~M.~Will, and K.~Young for many
useful discussions.  This work was supported by the U.~S.~National Science
Foundation via Grants  Nos.~PHY89--06286, PHY91--16682, PHY85--13953,
and PHY89--22140 at Washington University in St.~Louis, and No.~PHY91--05935
at the University of Wisconsin--Milwaukee.
\blankline
\hrule
\item{[1]}J.~A.~Wheeler, Ann.~Phys.~(NY) {\bf 2,} 604 (1957);
{\it Geometrodynamics\/} (Academic Press, New York, 1962), pp.~71--83.

\item{[2]}R.~P.~Geroch, J.~Math.~Phys.~{\bf 8,} 782 (1967); P.~Yodzis,
Commun.~Math.~Phys.~{\bf 26,} 39 (1972).

\item{[3]}For the numerous works on Euclidean quantum gravity and Euclidean
wormholes, see J.~J.~Halliwell, ``A Bibliography of Papers on Quantum
Cosmology,'' Int.~J.~Mod.~Phys.~{\bf A5,} 2473 (1990); ``Introductory
Lectures on Quantum Cosmology,'' in {\it Quantum Cosmology and Baby Universes:
Jerusalem Winter School for Theoretical Physics Vol.~7,} S.~Coleman,
J.~Hartle, T.~Piran, and S.~Weinberg, eds.~(World Scientific, Singapore, 1991),
pp.~159--243.

\item{[4]}See, e.g., W.~Unruh, Phys.~Rev.~D {\bf 40,} 1053 (1989) and
references cited therein.

\item{[5]}S.~W.~Hawking, Phys.~Rev.~D {\bf 37,} 904 (1988); S.~B.~Giddings
and A.~Strominger, Nucl.~Phys.~{\bf B307,} 857 (1988); S.~Coleman,
Nucl.~Phys.~{\bf B307,} 867 (1988); {\bf B310,} 643 (1988); also papers
in Ref.~[3].

\item{[6]}M.~S.~Morris, K.~S.~Thorne and U.~Yurtsever,
Phys.~Rev.~Lett.~{\bf 61,} 1446 (1988); V.~P.~Frolov and I.~D.~Novikov,
Phys.~Rev.~D {\bf 42,} 1057 (1990); J.~Friedman, M.~S.~Morris, I.~D.~Novikov,
F.~Echeverria, G.~Klinkhammer, K.~S.~Thorne, and U.~Yurtsever,
Phys.~Rev.~D {\bf 42,} 1915 (1990).

\item{[7]}M.~Visser, Phys.~Rev.~D {\bf 39,} 3182 (1989); {\bf 41,} 1116
(1990); Nucl.~Phys.~{\bf B328,} 203 (1989).

\item{[8]}W.~Israel, Nuov.~Cim.~{\bf 44B,} 1 (1966); {\bf 48B,} 463~(E) (1967);
C.~W.~Misner, K.~S.~Thorne, and J.~A.~Wheeler, {\it Gravitation\/} (Freeman,
San Francisco, 1973), pp.~551--555; S.~K.~Blau, E.~I.~Guendelman,
and A.~H.~Guth, Phys.~Rev.~D {\bf 35,} 1747 (1987).

\item{[9]}T.~A.~Roman, Phys.~Rev.~D {\bf 33,} 3526 (1986); {\bf 37,} 546
(1988).

\item{[10]}M.~Visser, Phys.~Lett.~{\bf B242,} 24 (1990); Phys.~Rev.~D {\bf 43,}
402 (1991).

\item{[11]}I.~H.~Redmount and W.-M.~Suen, Int.~J.~Mod.~Phys.~{\bf A,}
in press.

\item{[12]}L.~S.~Schulman, {\it Techniques and Applications of Path
Integration,} (Wiley, New York, 1981), pp.~92--96.

\item{[13]}I.~H.~Redmount and W.-M.~Suen, Phys.~Rev.~D, in preparation.

\item{[14]}J.~B.~Hartle, Phys.~Rev.~D {\bf 38,} 2985 (1988).

\item{[15]}L.~S.~Schulman, Ref.~12, p.~42--44.

\item{[16]}I.~H.~Redmount, W.-M.~Suen, and K.~Young, Phys.~Rev.~D, in
preparation.
\vfill\eject
\centerline{\sl Figure Caption}
\blankline
\blankline
FIG.~1.  Evolution of a wormhole wave function~$\psi(R,t)$, as effected
by the propagator~$G^{\rm (WKB)}$; the squared magnitude of~$\psi$ is shown.
The initial wave function used is
$\psi(R,0)=(2/\pi)^{1/4}\exp[-(R-10)^2]$.  All quantities are in Planck units.
\vfil\eject\end